\documentclass[10pt,twocolumn]{article}
\usepackage[margin=0.75in]{geometry}
\usepackage{amsmath,amssymb}
\usepackage{graphicx,booktabs,float}
\usepackage{pgfplots}
\pgfplotsset{compat=1.18}
\usepackage{tikz}
\usepackage[numbers,sort&compress]{natbib}
\usepackage{hyperref}
\usepackage{xcolor,microtype}
\usepackage{enumitem}
\setlist{nosep,leftmargin=*}

\title{\textsc{Instrumental}: Automatic Synthesizer Parameter\\Recovery from Audio via Evolutionary Optimization}
\usepackage{orcidlink}
\author{Philipp Bogdan\;\orcidlink{0009-0009-7405-3555}}
\date{}

\begin{document}
\maketitle

\begin{abstract}
\noindent Existing audio-to-MIDI tools extract notes but discard the timbral characteristics that define an instrument's identity. We present \textsc{Instrumental}, a system that recovers continuous synthesizer parameters from audio by coupling a differentiable 28-parameter subtractive synthesizer with CMA-ES, a derivative-free evolutionary optimizer. We optimize a composite perceptual loss combining mel-scaled STFT, spectral centroid, and MFCC divergence, achieving a matching loss of 2.09 on real recorded audio. We systematically evaluate eight hypotheses for improving convergence and find that only parametric EQ boosting yields meaningful improvement. Our results show that CMA-ES outperforms gradient descent on this non-convex landscape, that more parameters do not monotonically improve matching, and that spectral-analysis initialization accelerates convergence over random starts.
\end{abstract}

\section{Introduction}

Sound design is central to music production, yet recreating a sound heard in a recording remains laborious. Existing tools extract \emph{symbolic} information: source separators like Spleeter~\cite{hennequin2020spleeter} and Demucs~\cite{defossez2021hybrid} isolate stems; transcription models like MT3~\cite{gardner2022mt3} identify notes and instrument classes. These answer \emph{what notes are played} but not \emph{what synthesis parameters would reproduce this timbre}.

Major DAWs (Logic Pro, Ableton Live, FL Studio) and standalone tools (Melodyne, Basic Pitch, AnthemScore) output MIDI notes only -- pitch, onset, duration, velocity. Research tools such as InverSynth~\cite{barkan2019inversynth}, Sound2Synth~\cite{chen2022sound2synth}, and Syntheon target parameter recovery for specific synthesizers (FM or Vital), but are trained on synthetic data and do not generalize across synth architectures. The gap between instrument \emph{classification} (``this is a synthesizer lead'') and instrument \emph{recreation} (``these are the filter cutoff, oscillator mix, and envelope settings'') remains largely open.

Prior work on automatic synthesizer programming~\cite{barkan2019inversynth,chen2022sound2synth,hayes2024diffmoog,shin2025synthrl,hayes2025flowsynth} has tackled the inverse problem with neural networks and RL, typically targeting a single synthesizer architecture. We present \textsc{Instrumental}, a system that:
\begin{enumerate}
\item Provides an end-to-end pipeline from audio to playable synth patch;
\item Systematically compares optimization strategies (CMA-ES, gradient descent, RL);
\item Empirically tests eight hypotheses for improving convergence.
\end{enumerate}

\section{Method}

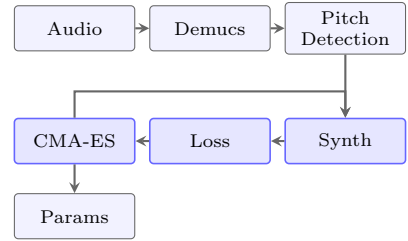
\begin{figure}[t]
\centering
\begin{tikzpicture}[
    node distance=0.9cm,
    box/.style={rectangle, draw=black!60, fill=blue!5, minimum width=1.6cm, minimum height=0.6cm, align=center, rounded corners=1.5pt, font=\scriptsize},
    loopbox/.style={rectangle, draw=blue!60, fill=blue!10, minimum width=1.6cm, minimum height=0.6cm, align=center, rounded corners=1.5pt, font=\scriptsize, line width=0.6pt},
    arr/.style={->, >=stealth, thick, black!60, font=\scriptsize},
]
\node[box] (audio) {Audio};
\node[box, right of=audio, node distance=1.8cm] (demucs) {Demucs};
\node[box, right of=demucs, node distance=1.8cm] (pitch) {Pitch\\Detection};
\node[loopbox, below of=pitch, node distance=1.5cm] (synth) {Synth};
\node[loopbox, left of=synth, node distance=1.8cm] (loss) {Loss};
\node[loopbox, left of=loss, node distance=1.8cm] (cmaes) {CMA-ES};
\node[box, below of=cmaes, node distance=1.0cm] (out) {Params};

\draw[arr] (audio) -- (demucs);
\draw[arr] (demucs) -- (pitch);
\draw[arr] (pitch) -- (synth);
\draw[arr] (synth) -- (loss);
\draw[arr] (loss) -- (cmaes);
\draw[arr] (cmaes.north) -- ++(0,0.35) -| (synth.north);
\draw[arr] (cmaes) -- (out);
\end{tikzpicture}
\caption{Pipeline: audio $\to$ Demucs separation $\to$ pitch detection $\to$ CMA-ES optimization loop $\to$ output parameters.}
\label{fig:pipeline}
\end{figure}

\textbf{Pipeline.} Input audio is separated into instrument stems via Demucs~\cite{defossez2021hybrid}, pitch-tracked with TorchCrepe~\cite{kim2018crepe}, segmented into individual notes with onset detection, then matched via CMA-ES optimization (Fig.~\ref{fig:pipeline}). The output is a parameter vector $\mathbf{p} \in [0,1]^{28}$ that can render the instrument at any pitch.

\subsection{Differentiable Subtractive Synthesizer}

We implement a fully differentiable subtractive synthesizer in PyTorch. The signal chain is:

\smallskip
\centerline{Oscillators $\to$ Mixer $\to$ LP Filter $\to$ EQ $\to$ Amp $\to$ Reverb}
\smallskip

\textbf{Oscillators.} Four waveform types: sawtooth ($2\phi/2\pi - 1$), pulse with variable duty cycle ($\tanh(20 \cdot (\sin\phi - \sin\phi_w))$ where $\phi_w = w \cdot 2\pi$), sine, and white noise. Each is mixed via learned weights. Up to 7 unison voices are generated per waveform, each detuned by $r_i = 2^{(d + s \cdot o_i)/12}$ where $d$ is global detune, $s$ is spread, and $o_i$ distributes voices symmetrically.

\textbf{Filter.} Rather than biquad coefficients (which introduce discontinuous gradients), we implement a frequency-domain filter with a sigmoid-shaped magnitude response:
\vspace{-1mm}
\begin{equation}
H(f) = \sigma\!\left(-\alpha\!\left(\frac{f}{f_c} - 1\right)\right) + Q \cdot 2\exp\!\left(-\frac{1}{2}\left(\frac{f/f_c - 1}{0.15}\right)^{\!2}\right)
\end{equation}
\vspace{-1mm}

\noindent where $f_c$ is cutoff, $\alpha \in [4,48]$ is the learnable slope, and $Q$ adds a resonant peak. Crucially, we add a \textbf{2-band parametric EQ} after the low-pass -- this allows the optimizer to \emph{boost} specific frequency bands, not only cut.

\textbf{Envelopes.} Two independent ADSR envelopes control amplitude and filter cutoff. A dedicated filter ADSR modulates the effective cutoff: $f_c^{\text{eff}} = f_c \cdot (1 + m \cdot (\bar{e}_{\text{filt}} - 0.5))$.

\textbf{Summary.} The 28 parameters span: oscillator mix (4), detune (1), filter (cutoff, resonance, slope: 3), filter ADSR (4), amplitude ADSR (4), parametric EQ (4: two bands $\times$ frequency + gain), pulse width (1), unison (voices, spread: 2), noise floor (1), reverb (size, mix: 2), filter envelope amount (1), and output gain (1).

\subsection{Loss Function}

We define a composite perceptual loss:
\vspace{-1mm}
\begin{equation}
\mathcal{L} = \mathcal{L}_{\text{mel}} + 0.1\,\mathcal{L}_{\text{cent}} + 0.05\,\mathcal{L}_{\text{mfcc}}
\end{equation}
\vspace{-1mm}

$\mathcal{L}_{\text{mel}}$ computes spectral convergence and log-magnitude L1 on mel-scaled spectrograms at three resolutions (FFT sizes 1024, 2048, 8192) with A-weighting via auraloss~\cite{steinmetz2020auraloss}. $\mathcal{L}_{\text{cent}}$ measures L1 distance between mean spectral centroids, normalized by the Nyquist frequency. $\mathcal{L}_{\text{mfcc}}$ is MSE over 13 MFCCs. The weights (1.0, 0.1, 0.05) were chosen so each term contributes roughly equally at convergence.

\subsection{Optimization}

\textbf{Spectral initialization.} Rather than random starting points, we analyze the target audio: spectral rolloff maps to filter cutoff, even-to-odd harmonic energy ratio determines oscillator type (high ratio $\to$ sawtooth, low $\to$ square), spectral flatness sets noise level, and RMS sets gain. This places the initial point in the correct basin.

\textbf{CMA-ES.} We use CMA-ES~\cite{hansen2016cmaes} with population $\lambda{=}40$, initial step size $\sigma_0{=}0.15$, bounds $[0,1]^{28}$, and a budget of $10^5$ evaluations. Each evaluation renders the synthesizer at all target pitches and computes $\mathcal{L}$.

\textbf{Multi-pitch fitting.} We optimize a single parameter vector across $K{=}3$ representative pitches simultaneously: $\mathcal{L}_{\text{total}}(\mathbf{p}) = \frac{1}{K}\sum_{k=1}^{K}\mathcal{L}(\text{render}(\mathbf{p}, f_0^{(k)}),\, x^{(k)})$. This prevents overfitting to one note and encourages the recovered patch to generalize across the instrument's range.

\textbf{Batched evaluation.} For throughput, we implement a fully batched synthesizer (all $\lambda$ candidates rendered as $(B, N)$ tensors in one call) and a batched loss function, achieving 553 evaluations/second on an Apple M4 with 10 CPU cores -- enabling 10K-evaluation experiments in 18 seconds.

\section{Experiments \& Results}

\textbf{Target.} A lead synthesizer extracted from a commercial song: 22 notes at three pitches (A3 = 221\,Hz, C\#4 = 278\,Hz, D4 = 295\,Hz), each ${\sim}150$\,ms, at ${\sim}100$\,BPM. Notes were segmented using librosa onset detection with $\delta{=}0.015$.

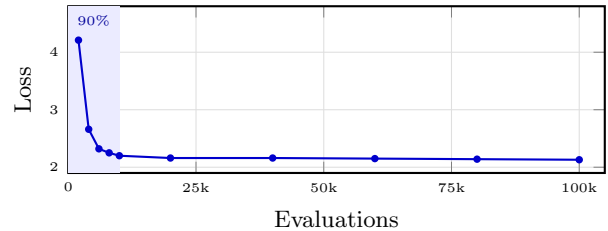
\begin{figure}[t]
\centering
\begin{tikzpicture}
\begin{axis}[
    width=\linewidth, height=3.8cm,
    xlabel={\small Evaluations}, ylabel={\small Loss},
    xmin=0, xmax=105000, ymin=1.9, ymax=4.8,
    xtick={0,25000,50000,75000,100000},
    xticklabels={0,25k,50k,75k,100k},
    ytick={2,3,4},
    scaled x ticks=false,
    xticklabel style={font=\tiny},
    yticklabel style={font=\tiny},
    xlabel style={font=\small}, ylabel style={font=\small},
    grid=major, grid style={gray!25}, thick,
]
\fill[blue!8] (axis cs:0,1.9) rectangle (axis cs:10000,4.8);
\node[font=\tiny, text=blue!60!black] at (axis cs:5000,4.55) {90\%};
\addplot[color=blue!80!black, mark=*, mark size=1pt, line width=0.8pt] coordinates {
    (2000,4.21) (4000,2.66) (6000,2.32) (8000,2.25) (10000,2.20)
    (20000,2.16) (40000,2.16) (60000,2.15) (80000,2.14) (100000,2.13)
};
\end{axis}
\end{tikzpicture}
\caption{CMA-ES convergence. 90\% of improvement occurs in the first 10K evaluations (shaded). The remaining 90K evals yield only 0.07 loss reduction.}
\label{fig:convergence}
\end{figure}

\subsection{Ablation: Synthesizer Parameters}

Table~\ref{tab:main} shows the progression as we incrementally add synthesizer modules. The largest single gain ($-49\%$) comes from adding unison voices and noise floor (15$\to$18 params), which allows the optimizer to model the ``thickness'' of the target sound. Adding pulse width and a learnable filter slope (18$\to$24) contributes a further 9\% reduction. The 2-band parametric EQ (24$\to$28) provides a modest but critical $-2\%$ gain -- the only modification that allows \emph{frequency boosting}.

At 29 parameters (adding unconstrained distortion drive, delay feedback, and vibrato depth), CMA-ES diverges: detune drifts to $+24$ semitones (two octaves), reverb mix reaches 43\%. The optimizer ``cheats'' the loss by exploiting extreme parameter values rather than matching the sound.

\begin{table}[t]
\centering
\small
\caption{Ablation over synthesizer parameters. Loss is the matching loss averaged over three target notes (lower is better).}
\label{tab:main}
\begin{tabular}{@{}clrc@{}}
\toprule
\textbf{Params} & \textbf{Modules added} & \textbf{Loss} & $\boldsymbol{\Delta}$ \\
\midrule
15 & Osc, filter, ADSR, reverb & 4.60 & -- \\
18 & + unison, noise floor & 2.34 & $-49\%$ \\
24 & + pulse width, filter slope & 2.13 & $-9\%$ \\
28 & + 2-band parametric EQ & \textbf{2.09} & $-2\%$ \\
29 & + distortion, delay, vibrato & \emph{div.} & -- \\
\bottomrule
\end{tabular}
\end{table}

\subsection{Hypothesis Testing}

We test eight modifications, each at 10K evaluations (Table~\ref{tab:hyp}). All tests start from the best 24-parameter solution (loss 2.27) and modify one component.

\textbf{Parametric EQ ($-8\%$).} The only positive result. The optimizer finds a $+1$\,dB Gaussian peak centered at 5.8\,kHz -- precisely the upper-mid ``presence'' region that music producers call ``zing.'' A low-pass filter, by definition, can only attenuate above the cutoff; the EQ adds the missing capability to boost.

\textbf{Chebyshev waveshaping ($+29\%$).} Five polynomial coefficients controlling individual harmonic amplitudes. Despite directly addressing the H3$>$H2 gap, the added parameters confused CMA-ES more than they helped -- the loss landscape becomes more multimodal with per-harmonic control.

\textbf{FM synthesis ($+17\%$).} Adding a 2-operator FM pair (ratio + modulation index). FM can produce H3$>$H2 spectra, but mixing FM with the existing subtractive chain at only 10K evaluations was insufficient for convergence.

\textbf{SPSA fine-tuning ($+11\%$).} Simultaneous Perturbation Stochastic Approximation after CMA-ES. The gradient estimates from SPSA were too noisy to improve on CMA-ES's already-converged solution.

\textbf{Multi-start CMA-ES ($+29\%$).} Eight independent runs with random perturbations. With only 1.25K evals per run, none reached the quality of a single focused 10K run.

\begin{table}[t]
\centering
\small
\caption{Hypothesis tests (10K evals each, baseline loss 2.27).}
\label{tab:hyp}
\begin{tabular}{@{}lr@{}}
\toprule
\textbf{Modification} & \textbf{$\Delta$ Loss} \\
\midrule
Chebyshev waveshaping & $+29\%$ \\
\textbf{Parametric EQ (2 bands)} & $\mathbf{-8\%}$ {\color{green!50!black}$\checkmark$} \\
TFS / envelope loss & $+17\%$ \\
SPSA fine-tuning & $+11\%$ \\
Multi-start CMA-ES ($8{\times}$) & $+29\%$ \\
FM synthesis (2-op) & $+17\%$ \\
\bottomrule
\end{tabular}
\end{table}

\subsection{Convergence and Spectral Analysis}

Fig.~\ref{fig:convergence} shows that 90\% of the total loss reduction occurs in the first 10K evaluations (18 seconds). The remaining 90K evaluations yield only $\sim$0.07 improvement, indicating an \emph{architectural floor} -- the limit of what this synthesizer topology can express.

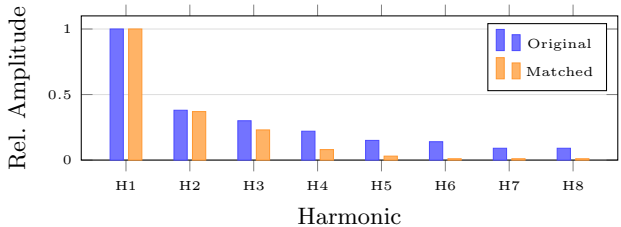
\begin{figure}[t]
\centering
\begin{tikzpicture}
\begin{axis}[
    width=\linewidth, height=3.5cm, ybar, bar width=5pt,
    xlabel={\small Harmonic}, ylabel={\small Rel.\ Amplitude},
    ymin=0, ymax=1.1, xtick=data,
    xticklabels={H1,H2,H3,H4,H5,H6,H7,H8},
    xticklabel style={font=\tiny},
    yticklabel style={font=\tiny},
    xlabel style={font=\small}, ylabel style={font=\small},
    ymajorgrids=true, grid style={gray!25},
    legend style={font=\tiny, at={(0.98,0.95)}, anchor=north east},
    enlarge x limits=0.1,
]
\addplot[fill=blue!55, draw=blue!70] coordinates {(1,1.0)(2,0.38)(3,0.30)(4,0.22)(5,0.15)(6,0.14)(7,0.09)(8,0.09)};
\addlegendentry{Original}
\addplot[fill=orange!60, draw=orange!80] coordinates {(1,1.0)(2,0.37)(3,0.23)(4,0.08)(5,0.03)(6,0.01)(7,0.01)(8,0.01)};
\addlegendentry{Matched}
\end{axis}
\end{tikzpicture}
\caption{Harmonic amplitudes: original vs.\ matched. H1--H3 are closely reproduced; H4--H8 diverge by 2--5$\times$, revealing the architectural floor of subtractive synthesis.}
\label{fig:harmonics}
\end{figure}

Fig.~\ref{fig:harmonics} reveals the source of this floor: upper harmonics (H4--H8) are 2--5$\times$ weaker in our output than in the original. Critically, the target exhibits H3 $>$ H2 (amplitudes 0.30 vs.\ 0.38) -- a harmonic profile impossible with standard sawtooth (where $\text{H}_n \propto 1/n$) or square waveforms. This strongly suggests the original sound uses FM synthesis, wavetable morphing, or waveshaping -- synthesis methods that our subtractive architecture cannot replicate.

\subsection{Optimizer Comparison}

\textbf{Adam gradient descent} plateaus at loss 4.60 (our 15-param starting point) due to local minima trapping -- the differentiable sigmoid filter provides clean gradients, but the non-convex oscillator-mix landscape defeats first-order methods.

\textbf{PPO reinforcement learning} (1M timesteps, 8 parallel environments, 69-dimensional observation space) learns to reduce loss from 4.16 to 2.31 in 50 steps -- matching CMA-ES -- but with 30 minutes of training versus CMA-ES's zero training cost. The reward signal ($\Delta$loss per step) remained at zero mean throughout training, indicating the need for better reward shaping.

\textbf{Pretrained encoder} (Syntheon, a neural network trained on Vital synthesizer presets) achieved loss 322 -- 138$\times$ worse than CMA-ES. This is expected: Syntheon was trained on Vital's architecture and parameter space, which differs from our custom synth. This result illustrates the cost of architecture mismatch, not a general limitation of neural approaches.

\section{Discussion}

\textbf{The EQ insight.} The single most impactful finding is that a standard subtractive low-pass filter creates an asymmetry: it can only \emph{remove} spectral energy. When the target sound has mid-high presence (1--5\,kHz) -- as most lead synthesizers do -- the optimizer is structurally unable to match it. Adding a 2-band parametric EQ with learnable center frequency and gain ($\pm 6$\,dB) breaks this asymmetry with just 4 parameters. The optimizer found a $+1$\,dB peak at 5.8\,kHz, exactly the ``presence'' range that defines the brightness of a lead sound.

\textbf{The parameter count paradox.} Naively, more parameters should mean more expressiveness. In practice, expanding from 24 to 29 unconstrained parameters caused CMA-ES to exploit physically implausible configurations. The lesson: each parameter must be \emph{constrained to a musically meaningful range}. When we constrained detune to $\pm 2$ semitones (from $\pm 24$) and fixed a tight $\sigma_0$, the 24-parameter synth converged reliably. Unconstrained parameters create deceptive minima that score well on the loss but sound wrong.

\textbf{Limitations.} We evaluate on a single target sound. The harmonic floor (H3 $>$ H2) indicates that subtractive synthesis cannot represent all timbres; FM, wavetable, or Chebyshev waveshaping oscillators would extend coverage. The loss plateau after 10K evaluations suggests that for further gains, one should either change the synth architecture or improve the loss function (e.g., CDPAM~\cite{manocha2021cdpam}, temporal fine structure losses).

\textbf{Future work.} Training a supervised encoder on synthetic (params, audio) pairs would amortize the optimization cost. Flow-matching generative models over the parameter posterior~\cite{hayes2025flowsynth} are particularly promising, as they naturally handle the many-to-one nature of synthesis (multiple parameter sets produce perceptually identical sounds). Richer synth architectures with wavetable oscillators and formant filters would raise the architectural ceiling.

{\small
\bibliographystyle{plainnat}
\bibliography{references}
}

\clearpage
\appendix
\section{Web Application}

We provide an interactive web application that wraps the full pipeline. Users upload audio, observe CMA-ES optimization in real time, and play the recovered synthesizer on an on-screen keyboard via WebAudio.

\begin{figure}[H]
\centering
\includegraphics[width=\linewidth]{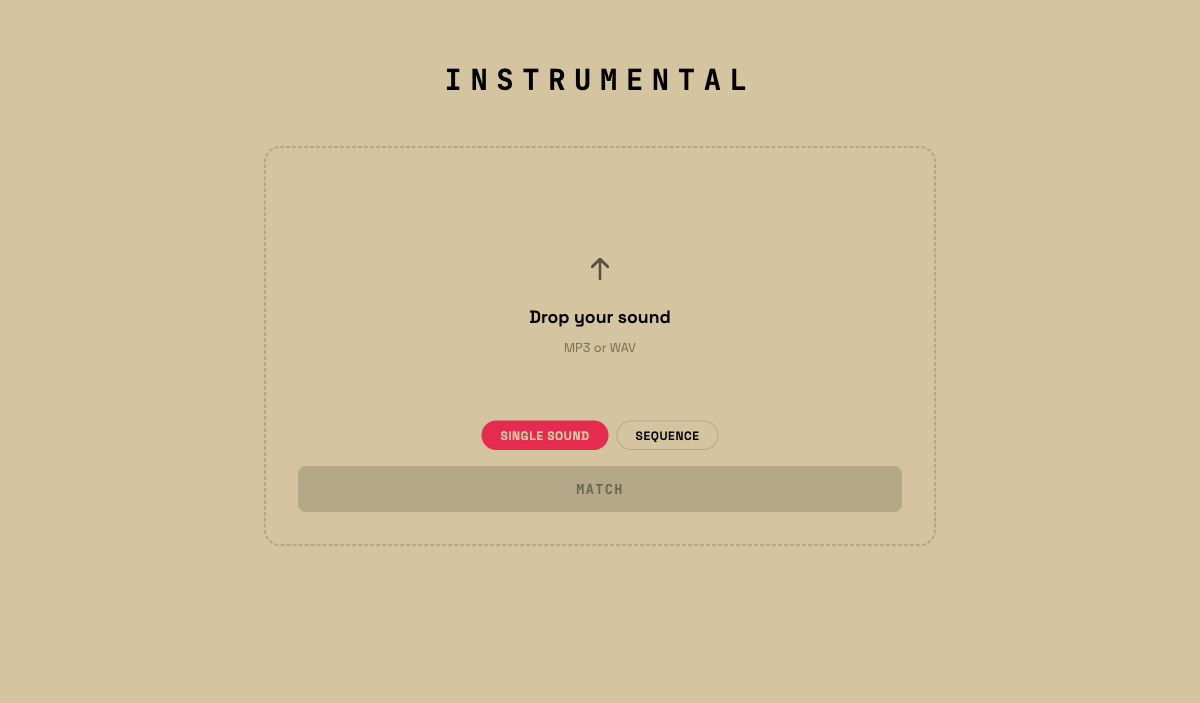}
\caption{Landing page. Users drop an audio file (MP3 or WAV) and select Single Sound or Sequence mode.}
\label{fig:app-landing}
\end{figure}

\begin{figure}[H]
\centering
\includegraphics[width=\linewidth]{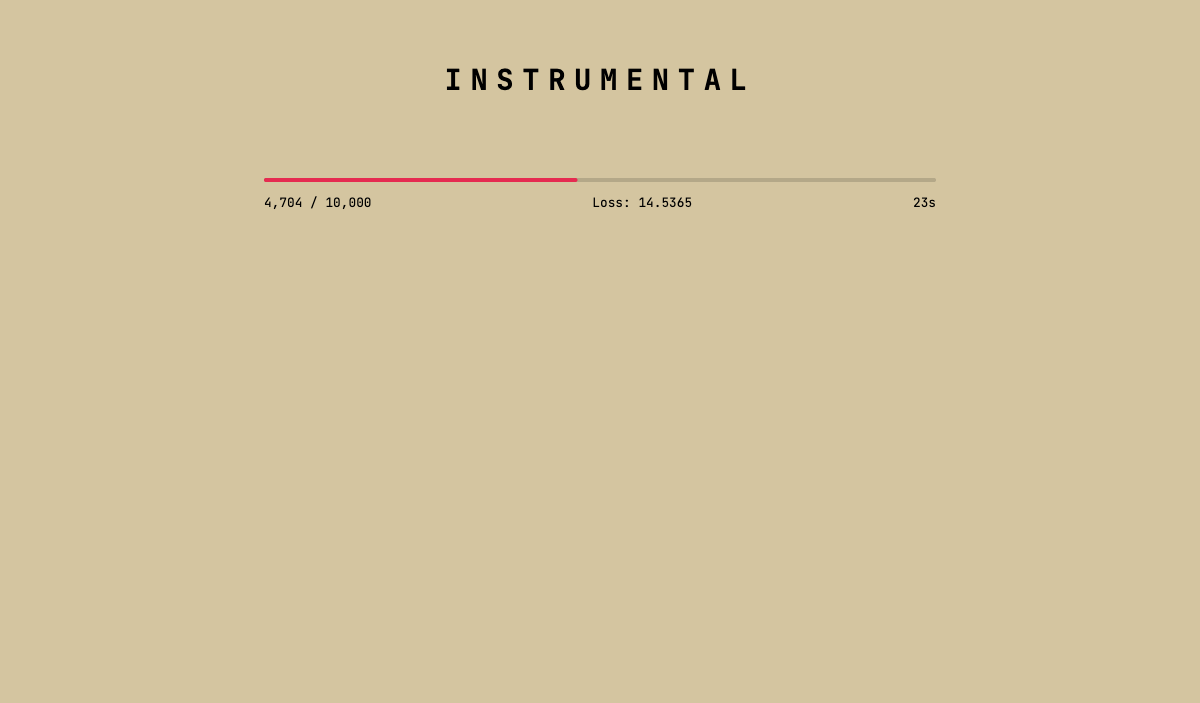}
\caption{Optimization in progress. The progress bar, evaluation count, current best loss, and elapsed time update live via WebSocket as CMA-ES runs across 7 CPU cores.}
\label{fig:app-progress}
\end{figure}

\begin{figure}[H]
\centering
\includegraphics[width=\linewidth]{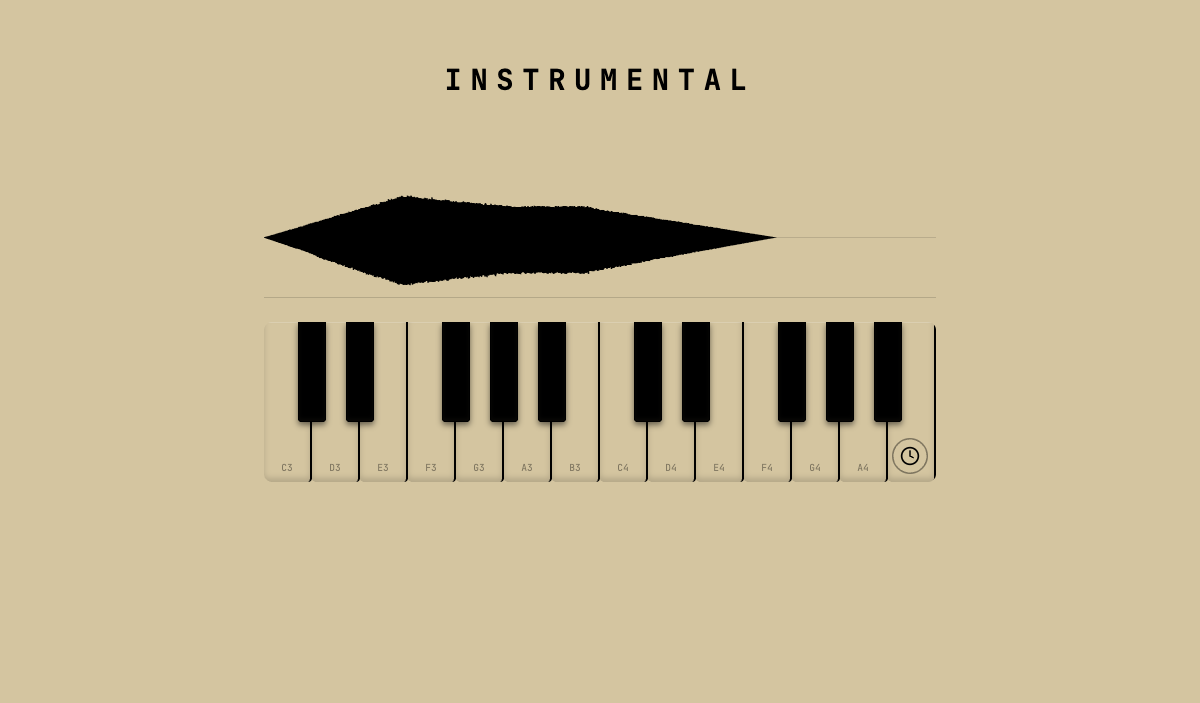}
\caption{Result view. The recovered waveform is displayed above a 2-octave piano keyboard. Each key plays the matched synthesizer at that pitch using WebAudio. The clock button (bottom-right) loads the best pre-computed preset (loss 2.09).}
\label{fig:app-keyboard}
\end{figure}

\end{document}